\documentstyle{article}
\begin{document}

\begin{center}
{\bf ONE-LOOP EFFECTIVE MULTI-GLUON LAGRANGIAN IN ARBITRARY DIMENSIONS} 
\vspace{0.5in}

E. RODULFO$^{*}$ and R. DELBOURGO$^{**}$ \vspace{0.1in}

{\small {\em Department of Physics, University of Tasmania\\G P O Box
252-21, Hobart 7001, Australia} \vspace{0.2in} }

{\small 25 September 1997 }
\end{center}

{\small \noindent
We exhibit the one-loop multi-gluon effective Lagrangian in any
dimension for a field theory with a quasilocal background, using the
background-field formalism. Specific results, including counter terms (up to
12 spacetime dimensions), have been derived, applied to the Yang-Mills theory
and found to be in agreement with other string-inspired approaches.}
\vspace{0.2in}

\noindent
{\bf 1. Introduction}
\vspace{.1in}

\noindent
{\normalsize The work presented in this paper has been motivated by a
reconsideration of photon-photon scattering, generalized to arbitrary
dimensional spacetime \cite{DR} and to any number of scattering photons.
Though an extremely feeble process and therefore very difficult to observe,
this nonlinear process of scattering photons by photons has attracted
considerable attention \cite{EK,KN1,JS,KN2,VO,SCH,FLI} since Euler first
considered its low-energy limit in 1935. In this paper, we generalize to a
non-Abelian analogue of this phenomenon: multi-gluon scattering in arbitrary
dimensional spacetime which is not as weak a process as its Abelian
counterpart, but is significantly richer in that gluons have classical
self-interaction; also the gluon number need no longer be even, unlike
photons. Multi-gluon scattering is mediated not only by meson and fermion
virtual particles, but by vector and ghost loops too, quite aside from the
tree-level contributions.}

{\normalsize A process involving an arbitrary number of gluons may be
represented by a sum over permutations of the Feynman diagram with those
many legs (including pinched diagrams). Clearly, direct evaluation of the
sum over Feynman amplitudes is very involved, so instead we shall attack
this problem in the background-field formalism \cite{BDW,GtH,GvNW,LA,IO,AGS}
where, if appropriate quasilocal conditions are assumed \cite{BD,DRM,HAG},
one can find exact expressions for the effective action without the tedious
computation of Feynman diagrams. We shall revisit this method in section 2
where we shall be led to an expression for the one-loop effective Lagrangian
for any gauge field theory with a quasilocal background in arbitrary
dimensions. We shall then extract from this Lagrangian specific results
including counterterms up to 12 spacetime dimensions and arrive at some
conclusions. (We have limited ourselves to $D \leq 12$ since higher numbers
of dimensions have not gained wide currency.) In section 3, we apply the
results of section 2 to the Yang-Mills theory with scalar bosons and Dirac
fermions in the hope that this will provide greater insight into the
structure of the effective Lagrangian associated with multi-gluon scattering
in arbitrary dimensions, including string models and M-theory. }

{\normalsize The requirement that the background be quasilocal (i.e.
possesses a covariantly constant field strength tensor) does not mean that
the results will be strictly valid only under that condition. In fact, as
will be demonstrated, general results not constrained by the quasilocal
assumption can be obtained for dimensions of practical importance, like $D
\leq 4$. This is due to the fact that the covariant derivative of the field
strength tensor of the background (which is set to zero in the quasilocal
case) plays a significant role only in six and higher dimensions \cite{vdV}. 
}

{\normalsize As a manifestation of the practical usefulness of working with
a quasilocal background, we have successfully recovered the pure Yang-Mills
Lagrangian (see eqn. 50) that portrays the curious disappearance of $%
F^{2}_{\mu \nu}$ one-loop divergences in 26 dimensions, as previously noted
in \cite{FT} in respect of dimensionally reduced supersymmetric theories
(and reproduced in reference \cite{MT} using open Bose string theory). It is
interesting to note how precisely the same results follow from seemingly
unrelated\footnote{The formal framework which relates string amplitudes to 
the background-field method has been explored fairly recently. See
for example \cite{ZBD,MS}.} approaches. They tell us that new renormalization
constants, involving higher powers of the curvature, must be introduced ab 
initio into quantum actions describing higher-dimensional gauge theories. What 
we and others have succeeded in doing is to calculate the (low-energy) 
one-loop contributions to such renormalization constants. In fewer space-time 
dimensions these terms of course represent finite calculable quantum corrections 
to the classical action.}

{\normalsize Although our original motivation focusses on a particular
process (multi-gluon scattering), the method and general results exhibited
in this paper may be applied to any ordinary renormalizable theory with an
appropriately chosen quasilocal background and seem readily extensible to
gravity \cite{CR,IO}. }
\vspace{0.2in}

\noindent
{\bf 2. The one-loop effective Lagrangian}
\vspace{.1in}

\noindent
{\normalsize The background-field procedure \cite{BDW,GtH,GvNW,LA,IO,AGS}
begins by replacing the field $A$ in the original classical Lagrangian $%
{\cal L}(A)$ by the sum $A+h $, where $A$ is now referred to as the
background (or external) field and $h$ the quantum (or internal) field.
Working in a Euclidean formulation, so there is no distinction between upper
and lower indices, the resulting Lagrangian is then expanded in $h$ as
follows. 
\begin{equation}
{\cal L} (A+h)\! = {\cal L} (A) + \frac{\partial {\cal L} (A)}{\partial A}h
+ \frac{1}{2} \partial_\mu h W_{\mu \nu} (A) \partial_{\nu} h + h N_{\mu}
(A) \partial_{\mu} h + \frac{1}{2} h M(A) h + {\cal O}(h^{3}),
\end{equation}
where $W$, $N$ and $M$ are external spacetime-dependent functions which may
be arranged to have the (anti)symmetry properties: 
\begin{eqnarray}
W_{\mu \nu}^{ij}&=&W_{\nu \mu}^{ij} = W_{\mu \nu}^{ji}  \nonumber \\
N_{\mu}^{ij}&=&-N_{\mu}^{ji} \\
M^{ij}&=&M^{ji},  \nonumber
\end{eqnarray}
by adding total spacetime derivatives to ${\cal L}$. The relevant Lagrangian
for loop diagrams is therefore given by \cite{GtH,GvNW} 
\begin{equation}
{\cal L}(A+h)-{\cal L}(A)-\frac{\partial {\cal L}(A)}{\partial A} = \frac{1}{%
2} \partial_{\mu}h W_{\mu \nu}(A) \partial_{\nu}h + h N_{\mu}(A)
\partial_{\mu}h + \frac{1}{2} h M(A) h + {\cal O}(h^{3}).
\end{equation}
One-particle-irreducible(1PI) loop diagrams are calculated by using the
quantum fields $h$ as internal lines, while the background fields $A$ appear
at external vertices. }

{\normalsize It then follows that one-loop quantum effects will be governed
only by terms bilinear in the quantum field $h$ and these are precisely the
terms explicitly written on the right-hand-side of (3). Except for the
theory of gravity which is not considered in this paper, we may safely
assume a flat metric, 
\begin{equation}
W_{\mu \nu}^{ij} = -\delta_{\mu \nu} \delta^{ij}.
\end{equation}
If one further defines the tensor quantities 
\begin{eqnarray}
X &\equiv& -m^{2}+M-N_{\mu}N_{\mu} \\
Y_{\mu \nu} &\equiv& \partial_{\mu}N_{\nu}-\partial_{\nu}N_{\mu} +
[N_{\mu},N_{\nu}]
\end{eqnarray}
which together with $h$ transform according to 
\begin{eqnarray}
X &\longrightarrow& e^{\Lambda(x)} X e^{-\Lambda(x)} \\
Y_{\mu \nu} &\longrightarrow& e^{\Lambda(x)} Y_{\mu \nu} e^{-\Lambda(x)} \\
h &\longrightarrow& e^{\Lambda(x)} h
\end{eqnarray}
for some arbitrary antisymmetric matrix $\Lambda^{ij}(x) $, then the
relevant bilinear Lagrangian may be cast in the manifestly gauge invariant
form 
\begin{equation}
L = h [(\partial_{\mu}+N_{\mu})^{2} + m^{2} + X] h/2.
\end{equation}
The generating function for connected Green functions associated with this
bilinear Lagrangian is given by 
\begin{equation}
\exp \frac{i}{\hbar} \int d^D\!x {\cal L}^{(1)} = \eta \int (dh) \exp \frac{i%
}{\hbar} \int d^D\!x \frac{1}{2} h [(\partial_{\mu}+N_{\mu})^2+m^{2}+X] h
\end{equation}
where ${\cal L}^{(1)}$ is the one-loop effective Lagrangian and the constant 
$\eta$ is chosen so that 
\begin{equation}
{\cal L}^{(1)} \stackrel{A \rightarrow 0}{\longrightarrow} 0.
\end{equation}
Differentiating (11) with respect to $X$ one finds that ${\cal L}^{(1)}$ is
determined by the 2-point Green function evaluated at the same point. 
\begin{equation}
\frac{\partial {\cal L}^{(1)}}{\partial X} = \frac{1}{2}\mbox{Tr} \langle
h(x)h(x) \rangle.
\end{equation}
Hence, one needs to solve the Green function equation associated with $L$
and this is given by 
\begin{equation}
[\partial^{2}+m^{2}+X(x)+\partial_{\mu}N_{\mu}(x)+2N_{\mu}(x)\partial_{\mu}+
N_{\mu}(x)N_{\mu}(x)]\frac{i}{\hbar}\langle
h(x)h(x^{\prime})\rangle=\delta^D(x,x^{\prime})
\end{equation}
But for an arbitrary background, this is clearly a nonlocal problem and one
is obliged to consider perturbative methods. Brown\ and Duff \cite{BD}
however, showed that by imposing appropriate restrictions on the background,
one may obtain ${\cal L}^{(1)}$ in (13) exactly. We shall closely follow
their method as we now assume a version of their quasilocal conditions on
the background field: 
\begin{eqnarray}
\partial_{\mu}Y_{\nu \rho} &=& [Y_{\nu \rho},N_{\mu}] \\
\partial_{\mu}X &=& [X,N_{\mu}]
\end{eqnarray}
It will be recognized that (15) is a non-Abelian analogue of the condition
imposed by Schwinger \cite{JS} on the Maxwell field strength tensor in
calculating one-loop effective Lagrangians for constant external
electromagnetic fields. This restriction accommodates a non-Abelian
background field with a covariantly constant field strength tensor \cite{AVR}%
. The complementary condition (16) is the simplest restriction that still
allows for non-Abelian gauge theories satisfying (15). It may be shown \cite
{BD} that a tensor $Y_{\mu \nu}$ that satisfies (15) possesses commuting
Lorentz components, or $[Y_{\mu \nu},Y_{\rho \sigma}]=0$. The field $N_{\mu}$
that satisfies (15) however, does not in general commute with itself (i.e., $%
[N_{\mu},N_{\nu}] \neq 0$). It is in this sense that a quasilocal background
specified by (15) and (16) may be taken non-Abelian. }

{\normalsize The most general form of $N_{\mu}$ that satisfies (15) is given
by \cite{BD} 
\begin{equation}
N_{\mu}(x) = - Y_{\mu \nu}(x) x_{\nu}/2 +
e^{\Lambda(x)}\partial_{\mu}e^{-\Lambda(x)}.
\end{equation}
Exploiting the gauge invariance of $L$, one can work in a gauge described by
the transformation \cite{BD} 
\begin{equation}
N_\mu \longrightarrow e^{-\Lambda(x)}(N_{\mu}+\partial_\mu) e^{\Lambda(x)}.
\end{equation}
This brings us to a gauge where $Y_{\mu \nu}$ is constant: $Y_{\mu \nu}(x) =
Y_{\mu \nu} (x^{\prime})$, whereupon 
\begin{equation}
N_{\mu} = -Y_{\mu \nu}(x^{\prime}) (x-x^{\prime})_\nu/2.
\end{equation}
It may be shown using (16) that in any gauge, $X$ commutes with $Y_{\mu \nu}$
and so in the gauge (18) $X$ is constant as well, i.e. $X(x) = X(x^{\prime})$%
. With these simplifications, the Green function equation (14) reduces to 
\begin{eqnarray}
[\partial^{2}+m^{2}+X(x^{\prime})&+&(x-x^{\prime})_{\mu}Y_{\mu
\nu}(x^{\prime})\partial_{\nu} -\frac{1}{4} (x-x^{\prime})_{\mu}Y^{2}_{\mu
\nu}(x^{\prime})(x-x^{\prime})_{\nu}]  \nonumber \\
& & \frac{i}{\hbar} \langle h(x)h(x^{\prime})\rangle =
\delta^D(x,x^{\prime}).
\end{eqnarray}
In order to solve (20), we substitute the trial solution 
\begin{equation}
\langle h(x)h(x^{\prime})\rangle=\hbar\int_0^{\infty}ds\int\frac{d^D\!p}{%
(2\pi i)^{D}} e^{-(m^{2}+X)s-P(s)-[i(x-x^{\prime})+Q(s)] \cdot p - p \cdot
R(s)\cdot p/2},
\end{equation}
where $P(s)$, $Q_{\mu}(s)$ and the symmetric nonsingular matrix $R_{\mu
\nu}(s)$ are to be determined in terms of $X$ and $Y$ subject to the
conditions ensuring consistency as the background field vanishes, 
\begin{eqnarray}
\lim_{A\rightarrow 0} \left[ 
\begin{array}{c}
P(s) \\ 
Q(s) \\ 
R(s)
\end{array}
\right] = \left[ 
\begin{array}{c}
0 \\ 
0 \\ 
-2s
\end{array}
\right].
\end{eqnarray}
$P$, $Q$ and $R$ are found to satisfy first order differential equations 
\cite{BD,HAG} which may readily be solved (see Appendix). Then (13) can be
integrated with respect to $X$ and the integration constant determined by
(12). The resulting one-loop effective Lagrangian may then be expressed as 
\cite{SCH} 
\begin{eqnarray}
{\cal L}^{(1)}&=&\pm\frac{\hbar}{2(4\pi)^{D/2}}\int_{0}^{\infty}ds%
\,s^{-1-D/2} e^{-m^{2}s}{\rm Tr}\sum_{p=o}^{\infty} \frac{(-1)^{p}}{p!} 
\nonumber \\
& &\left\{ \left[Xs+\sum_{n=1}^{\infty}\frac{2^{2n}B_{2n}}{4n(2n)!} {\rm tr}
(Ys)^{2n}\right]^{p} - [X(0)s]^p \right \}
\end{eqnarray}
where the overall sign is $(+)$ for bosons and $(-)$ for fermions. Tr
denotes a trace over internal indices (including possibly spinor indices)
while tr stands for a purely Lorentz trace. $X(0)$ is $X$ evaluated at zero
background and $B_{n}$ are Bernoulli numbers. The result (23) which may be
continued to arbitrary $D$, summarizes the contributions from all one-loop
Feynman diagrams possessing arbitrarily many legs. It should also be noted
that (23) remain valid in strong (but quasilocal) fields \cite{AVR} as it
contains all orders of the coupling constant which enter the expression
through $X$ and $Y$. }

{\normalsize Let us now extract some results from the one-loop effective
Lagrangian (23). First, let us denote by ${\cal L}^{(1)}(D,p)$ the part of $%
{\cal L}^{(1)}$ which is of order $q$ in $X$ and order $r$ in $Y$ such that $%
q+r=p$, i.e. 
\begin{equation}
{\cal L}^{(1)}(D,p) \sim \sum_{q+r=p} X^{q} Y^{r}.
\end{equation}
The first six non-vanishing ${\cal L}^{(1)} (D,p)$'s are 
\begin{equation}
{\cal L}^{(1)}(D,1) = \frac{\hbar}{2(4\pi)^{D/2}}\frac{\Gamma(1-D/2)}{m^{2-D}%
} {\rm Tr}\{-[X-X(0)]\}
\end{equation}
\begin{equation}
{\cal L}^{(1)}(D,2) = \frac{\hbar}{2(4\pi)^{D/2}}\frac{\Gamma(2-D/2)}{m^{4-D}%
} {\rm Tr}\left\{\frac{1}{2}[X^{2}-X(0)^{2}]-\frac{1}{12}{\rm tr}%
Y^{2}\right\}
\end{equation}
\begin{equation}
{\cal L}^{(1)}(D,3) = \frac{\hbar}{2(4\pi)^{D/2}}\frac{\Gamma(3-D/2)}{m^{6-D}%
} {\rm Tr}\left\{-\frac{1}{6}[X^{3}-X(0)^{3}]+\frac{1}{12}X{\rm tr}%
Y^{2}\right\}
\end{equation}
\begin{eqnarray}
{\cal L}^{(1)}(D,4)&=&\frac{\hbar}{2(4\pi)^{D/2}}\frac{\Gamma(4-D/2)}{m^{8-D}%
} {\rm Tr} \left\{\frac{1}{24}[X^{4}-X(0)^{4}]-\frac{1}{24}X^{2}{\rm tr}%
Y^{2} \right.  \nonumber \\
& &\left.+\frac{1}{288}({\rm tr}Y^{2})^{2}+\frac{1}{360}{\rm tr}Y^{4}\right\}
\end{eqnarray}
\begin{eqnarray}
{\cal L}^{(1)}(D,5)&=&\frac{\hbar}{2(4\pi)^{D/2}}\frac{\Gamma(5-D/2)}{%
m^{10-D}} {\rm Tr}\left\{-\frac{1}{120}[X^5-X(0)^5]+\frac{1}{72}X^3{\rm tr}%
Y^2  \right.  \nonumber \\
& & \left.- \frac{1}{288}X({\rm tr}Y^2)^2-\frac{1}{360}X{\rm tr}Y^4 \right\}
\end{eqnarray}
\[
{\cal L}^{(1)}(D,6) = \frac{\hbar}{2(4\pi)^{D/2}} \frac{\Gamma(6-D/2)}{%
m^{12-D}}{\rm Tr}\left\{\frac{1}{720}[X^6-X(0)^6]- \frac{1}{288}X^{4}{\rm tr}%
Y^{2} + \right. 
\]
\begin{equation}
\left. \frac{1}{576}X^2({\rm tr}Y^2)^2+\frac{1}{720}X^2{\rm tr}Y^4 \!-\frac{%
1}{10368}({\rm tr}Y^2)^3 -\!\frac{1}{4320}{\rm tr}Y^2 {\rm tr}Y^4 - \frac{1}{%
5670}{\rm tr}Y^6 \right\}.
\end{equation}
}

{\normalsize The divergent part of ${\cal L}^{(1)} (D,p)$ is given by ${\cal %
L}^{(1)} (D \rightarrow 2p ,p)$. For the results (25) to (30), the
corresponding divergent parts (as $D$ approaches an integer value $2p$) are
picked out as 
\begin{equation}
{\cal L}^{(1)}(D\rightarrow 2,1)=\frac{\hbar}{4\pi(2-D)}{\rm Tr}\{-[X-X(0)]\}
\end{equation}
\begin{equation}
{\cal L}^{(1)}(D \rightarrow 4,2) = \frac{\hbar}{16\pi^{2}(4-D)} {\rm Tr}
\left\{\frac{1}{2}[X^{2}-X(0)^{2}]-\frac{1}{12}{\rm tr}Y^{2} \right\}
\end{equation}
\begin{equation}
{\cal L}^{(1)}(D \rightarrow 6,3) = \frac{\hbar}{64\pi^{3}(6-D)} {\rm Tr}
\left\{-\frac{1}{6}[X^{3}-X(0)^{3}]+\frac{1}{12}X{\rm tr}Y^{2}\right\}
\end{equation}
\begin{eqnarray}
&{\cal L}^{(1)}(D \rightarrow 8,4)=& \frac{\hbar}{256\pi^{4}(8-D)}{\rm Tr}
\left\{\frac{1}{24}[X^{4}-X(0)^{4}]-\frac{1}{24}X^{2}{\rm tr}Y^{2} \right. 
\nonumber \\
& & \left.+\frac{1}{288}({\rm tr}Y^2)^2+\frac{1}{360}{\rm tr}Y^{4} \right\}
\end{eqnarray}
\[
{\cal L}^{(1)}(D \rightarrow 10,5) =\frac{\hbar}{1024\pi^{5}(10-D)}{\rm Tr}
\left\{-\frac{1}{120}[X^{5}-X(0)^{5}]+\frac{1}{72}X^{3}{\rm tr}Y^{2}
\right. 
\]
\begin{equation}
\left.-\frac{1}{288}X({\rm tr}Y^{2})^2-\frac{1}{360}X{\rm tr}Y^4 \right\}
\end{equation}
\[
{\cal L}^{(1)}(D\rightarrow 12,6) =\frac{\hbar}{4096\pi^{6}(12-D)}{\rm Tr}
\left\{\frac{1}{720}[X^{6}-X(0)^{6}]-\frac{1}{288}X^{4}{\rm tr}Y^{2}+
\right. 
\]
\begin{equation}
\left.\frac{1}{576}X^{2}({\rm tr}Y^{2})^2+\frac{1}{720}X^{2}{\rm tr}Y^4 -%
\frac{1}{10368}({\rm tr}Y^{2})^{3}-\frac{1}{4320}{\rm tr}Y^{2} {\rm tr}Y^{4}-%
\frac{1}{5670}{\rm tr}Y^{6} \right\}.
\end{equation}
Any higher dimensional Lagrangians are probably irrelevant to the most
popular physical models. }

{\normalsize One-loop counterterms are usually defined as the negative of
our ${\cal L}^{(1)} (D \rightarrow 2 p ,p)$. The results (31) and (32)
supply the divergent Lagrangians in $D=2$ and $D=4$, respectively. These
results are valid in general (even in the nonquasilocal case) as may be
checked with references \cite{FT,vdV,GtH,IO} because the covariant
derivatives ${\cal D}Y$ (where ${\cal D}_{\mu} \equiv \partial_{\mu} +
[N_{\mu}, ]$), only begin to appear in the invariants for $D \geq 6$ \cite
{vdV}. However, the results (33) to (36) for $D=6,8,10,12$ are valid only in
the quasilocal case. The results for $D=6,8,10$ may still be compared with
references\cite{FT,vdV} provided the quasilocal conditions (15) and (16) are
imposed on their results. The presence of covariant derivatives however,
provides some degree of arbitrariness in the choice of invariants due to the
Bianchi identities and possible partial integrations on the covariant
derivatives. For instance, the invariant $Y_{\mu \nu} Y_{\nu \rho} Y_{\rho
\mu}$ appears in the results of \cite{FT,vdV} but does not appear in the
quasilocal case (33) because the Bianchi identities allow us to write (up to
a total divergence) 
\begin{equation}
Y_{\mu \nu}Y_{\nu\rho}Y_{\rho\mu}=\frac{1}{2}({\cal D}_{\mu}Y_{\mu \nu})^{2}
+\frac{1}{2} ({\cal D}_{\mu} Y_{\nu \rho})({\cal D}_{\nu} Y_{\rho \mu}),
\end{equation}
which clearly vanishes in the quasilocal case. Our results for $D \geq 6$
may be viewed as unique quasilocal limiting expressions of more general
frameworks since all arbitrariness in the choice of invariants disappears in
this limit. }

{\normalsize One may also indirectly compare the results of this section
with those of Avramidi \cite{AVR}, who calculated the asymptotic
coefficients of the heat kernel up to $D=16$ essentially, although some work is 
needed before the results can be immediately compared. }
\vspace{.2in}

\noindent
{\bf 3. Yang-Mills Theory}
\vspace{.1in}

\noindent
{\normalsize In order to apply the results of section 2 to a specific field
theory, one needs to determine for the theory in question the relevant
second order operator 
\begin{equation}
\Delta \equiv (\partial_{\mu} + N_{\mu})^{2} + m^{2} + X
\end{equation}
which appears in the bilinear Lagrangian (10) and from this identify $N_{\mu}
$, $m^{2}$ and $X$. $Y_{\mu \nu}$ follows immediately from $N_{\mu}$ through
(6). Once these expressions are correctly identified, the results of section
2 can be easily applied. As will be demonstrated in this section in the
case of Yang-Mills theory, the term $m^{2}+X$ in (38) plays the role of a
generic `source' which determines the type of virtual particle loop that
mediates the interaction. We begin by considering the pure Yang-Mills theory
in section 3.1. Scalar bosons and Dirac fermions in a Yang-Mills background
are discussed in sections 3.2 and 3.3. Results for the Yang-Mills theory
incorporating the mesons and fermions close this section. }
\vspace{.2in}

\noindent
{\small \bf 3.1 Pure Yang-Mills theory}
\vspace{.1in}

\noindent
{\normalsize Performing the background-field replacement $A_{\mu}
\rightarrow A_{\mu}+a_{\mu}$ in the bare Lagrangian for the pure Yang-Mills
theory \cite{GtH,DRM} 
\begin{equation}
{\cal L} = -\frac{1}{4} F^{a}_{\mu \nu} F^{a}_{\mu \nu}
\end{equation}
where 
\begin{equation}
F^{a}_{\mu \nu} = \partial_{\mu}A^{a}_{\nu}-\partial_{\nu}A^{a}_{\mu}+
gf^{abc}A^{b}_{\mu}A^{c}_{\nu}
\end{equation}
and using the Feynman-background gauge \cite{GtH,DRM,LA} 
\begin{equation}
{\cal L}_{fix} = -\frac{1}{2}[(\partial_{\mu}\delta^{ac}+
gf^{abc}A^{b}_{\mu})a^{c}_{\mu}]^{2}
\end{equation}
one finds that the relevant bilinear Lagrangian is 
\begin{equation}
2L_{pYM}=a^{a}_{\alpha}[(\partial_{\mu}\delta^{ab} \delta_{\alpha
\beta}-gf^{abc}A^{c}_{\mu}\delta_{\alpha \beta})^{2}- 2gf^{abc}F^{c}_{\alpha
\beta}]a^{b}_{\beta} + \eta^{a}_{i}
(\partial_{\mu}\delta^{ab}-gf^{abc}A^{c}_{\mu})^{2}\eta^{b}_{i}
\end{equation}
The first term in (42) gives the vector part while the second gives the
contribution of the two fictitious fields $\eta_{i}$, $i=1,2$. Hence, the
relevant second order vector and ghost operators are, respectively 
\begin{equation}
(\Delta_{vector})^{ab}_{\alpha \beta}=(\partial_{\mu}\delta^{ab}
\delta_{\alpha \beta}-gf^{abc}A^{c}_{\mu}\delta_{\alpha \beta})^{2}-
2gf^{abc}F^{c}_{\alpha \beta}
\end{equation}
\begin{equation}
(\Delta_{ghost})^{ab}=(\partial_{\mu}\delta^{ab}-gf^{abc}A^{c}_{\mu})^2
\end{equation}
Comparison with (38) immediately yields: 
\begin{equation}
(N_{vector})^{ab}_{\mu \alpha \beta}=-gf^{abc}A^{c}_{\mu}\delta_{\alpha
\beta}
\end{equation}
\begin{equation}
(Y_{vector})^{ab}_{\mu\nu\alpha\beta}=-gf^{abc}F^c_{\mu\nu}\delta_{\alpha%
\beta}
\end{equation}
\begin{equation}
(X_{vector})^{ab}_{\alpha \beta}=-2gf^{abc}F^{c}_{\alpha \beta}
\end{equation}
\begin{equation}
(N_{ghost})^{ab}_{\mu}=-gf^{abc}A^{c}_{\mu}
\end{equation}
\begin{equation}
(Y_{ghost})^{ab}_{\mu \nu}=-gf^{abc}F^{c}_{\mu \nu}
\end{equation}
together with the vanishing expressions $m_{vector}=m_{ghost}=X_{ghost}=0$.
Note that the ghost effective Lagrangian acquires an overall factor of $-2$
resulting from the two ``fermionic'' fields $\eta_{1}$ and $\eta_{2}$. }

{\normalsize Using the results of section 2 one finds that the first few
nonvanishing ${\cal L}^{(1)}_{pYM} (D,p)$'s are 
\begin{equation}
{\cal L}^{(1)}_{pYM}(D,2)=\frac{\hbar g^{2}}{2(4\pi)^{D/2}} \left(\frac{26-D%
}{12}\right)\int_{0}^{\infty}\,ds\,s^{1-D/2} \,CF^{a}_{\mu\nu} F^a_{\mu\nu}
\end{equation}
\begin{eqnarray}
&{\cal L}^{(1)}_{pYM}(D,4)&=\frac{\hbar g^{4}}{2(4\pi)^{D/2}}\int_0^{\infty}
ds\,s^{3-D/2} \,f^{abe}f^{bcf}f^{cdg}f^{dah}  \nonumber \\
& &\left\{ \frac{238+D}{360}F^{e}_{\mu\nu}F^f_{\nu \rho}F^{g}_{\rho\sigma}
F^{h}_{\sigma \mu}+\frac{-50+D}{288}F^{e}_{\mu \nu} F^{f}_{\nu \mu}
F^{g}_{\rho \sigma}F^{h}_{\sigma \rho} \right\}
\end{eqnarray}
\begin{eqnarray}
{\cal L}^{(1)}_{pYM}(D,5)&=&-\frac{\hbar g^5}{2(4\pi)^{D/2}}\int_0^{\infty}
ds\,s^{4-D/2} \,f^{abf}f^{bcg}f^{cdh}f^{dei}f^{eaj}\cdot  \nonumber \\
& & \frac{4}{15}F^{f}_{\mu\nu} F^{g}_{\nu \rho}F^{h}_{\rho
\sigma}F^{i}_{\sigma \lambda}F^j_{\lambda \mu}
\end{eqnarray}
\begin{eqnarray}
& {\cal L}^{(1)}_{pYM}(D,6)=\frac{\hbar g^{6}}{2(4\pi)^{D/2}}\int_0^{\infty}
ds\,s^{5-D/2} \,f^{abg}f^{bch}f^{cdi}f^{dej}f^{efk}f^{fal}  \nonumber \\
& \left\{\frac{506-D}{5670}F^{g}_{\mu\nu}F^{h}_{\nu \rho}F^{i}_{\rho\sigma}
F^j_{\sigma \lambda}F^{k}_{\lambda \tau}F^l_{\tau \mu} -\frac{214+D}{4320}%
F^{g}_{\mu \nu}F^h_{\nu\mu}F^i_{\rho \sigma} F^j_{\sigma
\lambda}F^k_{\lambda \tau}F^l_{\tau \rho}+\right.  \nonumber \\
& \left.\frac{74-D}{10368}F^g_{\mu\nu} F^h_{\nu\mu}F^i_{\rho\sigma}
F^j_{\sigma\rho} F^k_{\lambda \tau}F^l_{\tau \lambda} \right\}
\end{eqnarray}
where $C \delta^{cd} = f^{abc} f^{abd}$ is the Casimir of the adjoint
representation. The result (50) clearly exhibits the curious absence of $%
F^2_{\mu \nu}$ one-loop divergences in the pure Yang-Mills theory in 26
spacetime dimensions, as noted in references \cite{FT,MT}. The divergent
part of (50) is the well known result \cite{GtH,GW,HP} 
\begin{equation}
{\cal L}^{(1)}_{pYM}(D \rightarrow 4,2)=\frac{\hbar g^{2}}{32\pi^{2}(4-D)} 
\frac{11}{3}\,CF^{a}_{\mu \nu} F^{a}_{\mu \nu}
\end{equation}
which is of course valid even in the nonquasilocal case. The divergent part
of the Lagrangian in 8 dimensions follows immediately from (51): 
\begin{eqnarray}
{\cal L}^{(1)}_{pYM}(D\rightarrow 8,4)&=&\frac{\hbar g^4}{256\pi^4(8-D)}
f^{abe}f^{bcf}f^{cdg}f^{dah}  \nonumber \\
& &\left(\frac{41}{60}F^e_{\mu \nu} F^f_{\nu \rho} F^g_{\rho
\sigma}F^{h}_{\sigma \mu}-\frac{7}{48}F^{e}_{\mu \nu} F^f_{\nu \mu}F^g_{\rho
\sigma}F^h_{\sigma \rho} \right)
\end{eqnarray}
(The quasilocal results (51) and (55) may be compared with equation (2.9) of
reference \cite{MT}.) For completeness, let us also write down the
quasilocal divergent Lagrangians in ten and twelve dimensions: 
\begin{equation}
{\cal L}^{(1)}_{pYM}(D\rightarrow 10,5)=-\frac{\hbar g^5f^{abf}
f^{bcg}f^{cdh}f^{dei}f^{eaj}}{1024\pi^{5}(10-D)}\frac{4}{15} F^{f}_{\mu \nu}
F^{g}_{\nu \rho}F^{h}_{\rho \sigma}F^{i}_{\sigma \lambda} F^{j}_{\lambda \mu}
\end{equation}
\[
{\cal L}^{(1)}_{pYM}(D\rightarrow\! 12,6) \!=\frac{\hbar g^6
f^{abg}f^{bch}f^{cdi}f^{dej}f^{efk}f^{fal}}{4096\pi^{6}(12-D)} \left(\frac{%
247}{2835}  F^{g}_{\mu \nu} F^{h}_{\nu \rho}F^{i}_{\rho \sigma}F^{j}_{\sigma
\lambda}  F^{k}_{\lambda \tau}F^{l}_{\tau \mu} \right. 
\]
\begin{equation}
\left.-\!\frac{113}{2160}F^{g}_{\mu \nu} F^{h}_{\nu\mu}F^{i}_{\rho \sigma}
F^{j}_{\sigma \lambda}F^{k}_{\lambda \tau}F^{l}_{\tau \rho}+\! \frac{31}{5184%
}F^{g}_{\mu \nu} F^{h}_{\nu\mu}F^{i}_{\rho \sigma} F^{j}_{\sigma
\rho}F^{k}_{\lambda \tau}F^{l}_{\tau \lambda} \right)
\end{equation}
}

\vspace{.2in}

\noindent
{\small \bf 3.2 Scalar bosons in a Yang-Mills background}
\vspace{.1in}

\noindent
{\normalsize The Lagrangian for scalar bosons in a classical background
gauge field $A_{\mu}$ takes the bilinear form \cite{MS} 
\begin{equation}
L_{scalar}=\phi^{\dagger}[(\partial_{\mu}-igT^{a}_{s}A^{a}_{\mu})^{2}+m^{2}]%
\phi
\end{equation}
where we normalize the generators according to 
\begin{equation}
{\rm Tr}\left(T^{a}_{s}T^{b}_{s}\right)=T_{s}\delta^{ab}.
\end{equation}
The relevant second order operator (38) for this theory is therefore 
\begin{equation}
\Delta_{scalar}=(\partial_{\mu}-igT^{a}_{s}A^{a}_{\mu})^{2}+m^{2}
\end{equation}
and one immediately identifies: 
\begin{equation}
(N_{scalar})_{\mu}=-igT^{a}_{s}A^{a}_{\mu}
\end{equation}
\begin{equation}
(Y_{scalar})_{\mu\nu}=-igT^{a}_{s}F^{a}_{\mu\nu}
\end{equation}
\begin{equation}
X_{scalar}=0.
\end{equation}
The results of the previous section now allow us to list down the first few
nonvanishing ${\cal L}^{(1)}_{scalar} (D,p)$'s: 
\begin{equation}
{\cal L}_{scalar}^{(1)}(D,2) = \frac{\hbar g^2}{2(4\pi)^{D/2}} \frac{%
\Gamma(2-D/2)}{m^{4-D}} \frac{1}{12} T_{s}F^{a}_{\mu\nu}F^{a}_{\mu\nu}
\end{equation}
\begin{eqnarray}
{\cal L}_{scalar}^{(1)}(D,4)&=&\frac{\hbar g^4}{2(4\pi)^{D/2}}
\frac{\Gamma(4-D/2)}{m^{8-D}}{\rm Tr} 
\left(T^a_sT^b_sT^c_sT^d_s \right)  \nonumber \\
& &\left\{ \frac{1}{360}F^{a}_{\mu \nu} F^b_{\nu \rho}F^c_{\rho \sigma}
F^d_{\sigma \mu}+\frac{1}{288}F^a_{\mu\nu}F^b_{\nu \mu}F^c_{\rho \sigma}
F^d_{\sigma \rho} \right\},
\end{eqnarray}
\begin{eqnarray}
&{\cal L}_{scalar}^{(1)}(D,6)=\frac{\hbar g^6}{2(4\pi)^{D/2}}\frac{\Gamma(6-D/2)}
{m^{12-D}}{\rm Tr} \left(T^a_sT^b_sT^c_sT^d_sT^e_sT^f_s \right)  \nonumber \\
& \left\{ \frac{1}{5670}F^{a}_{\mu\nu} F^b_{\nu\rho}F^c_{\rho \sigma}
F^d_{\sigma \lambda}F^e_{\lambda\tau}F^f_{\tau\mu} +\frac{1}{4320}
F^a_{\mu\nu}F^b_{\nu\mu}F^c_{\rho\sigma} F^d_{\sigma\lambda}F^e_{\lambda
\tau}F^f_{\tau\rho} \right.  \nonumber \\
& \left. + \frac{1}{10368}F^a_{\mu \nu}F^b_{\nu\mu}F^c_{\rho\sigma}
F^d_{\sigma \rho}F^e_{\lambda \tau}F^f_{\tau \lambda} \right\}.
\end{eqnarray}
The corresponding divergent Lagrangians are 
\begin{equation}
{\cal L}_{scalar}^{(1)}(D\rightarrow 4,2) = \frac{\hbar g^2}{32\pi^{2}(4-D)} 
\frac{1}{6} T_{s}F^{a}_{\mu\nu}F^{a}_{\mu\nu}
\end{equation}
\begin{eqnarray}
{\cal L}_{scalar}^{(1)}(D\rightarrow 8,4)&=&\frac{\hbar g^4}{256\pi^{4}(8-D)}
{\rm Tr}\left(T^{a}_{s}T^{b}_{s}T^{c}_{s}T^{d}_{s} \right)  \nonumber \\
& & \left\{ \frac{1}{360} F^{a}_{\mu \nu} F^{b}_{\nu \rho}F^{c}_{\rho
\sigma}F^{d}_{\sigma \mu}+ \frac{1}{288}F^{a}_{\mu \nu} F^{b}_{\nu
\mu}F^{c}_{\rho \sigma} F^{d}_{\sigma \rho} \right\}
\end{eqnarray}
\begin{eqnarray}
&{\cal L}_{scalar}^{(1)}(D\rightarrow 12,6)=\frac{\hbar g^6}{4096\pi^{6}(12-D)} 
{\rm Tr}\left(T^a_sT^b_sT^c_sT^d_sT^e_sT^f_s \right)  \nonumber \\
& \left\{ \frac{1}{5670}F^a_{\mu\nu}F^b_{\nu\rho}F^c_{\rho\sigma} F^d_{\sigma
\lambda}F^e_{\lambda\tau}F^f_{\tau\mu} +\frac{1}{4320}F^a_{\mu\nu}F^b_{\nu
\mu}F^c_{\rho\sigma}F^d_{\sigma\lambda} F^e_{\lambda\tau}F^f_{\tau\rho}
\right.  \nonumber \\
& \left. +\frac{1}{10368}F^a_{\mu\nu}F^b_{\nu\mu}F^c_{\rho\sigma}
F^d_{\sigma\rho} F^e_{\lambda\tau}F^f_{\tau \lambda} \right\}.
\end{eqnarray}
}
\vspace{.2in}

\noindent
{\small \bf 3.3 Dirac fermions in a Yang-Mills background}
\vspace{.1in}

\noindent
{\normalsize The Lagrangian for Dirac fermions in a classical background
gauge field $A_{\mu}$ is usually written in a form that involves a first
order operator 
\begin{equation}
L_{fermion}=\psi^{\dagger}[i\gamma_{\mu}(\partial_{\mu}-igT^a_{f}A^a_{\mu})
-m{\bf 1}]\psi
\end{equation}
where $\gamma_{\mu}$ represents a $2^{[D/2]} \times 2^{[D/2]}$ Dirac matrix
normalized as usual by 
\begin{equation}
\{\gamma_{\mu},\gamma_{\nu}\}=2\delta_{\mu\nu}{\bf 1}
\end{equation}
\begin{equation}
\sigma_{\mu\nu}=\frac{i}{2}[\gamma_{\mu},\gamma_{\nu}]
\end{equation}
\begin{equation}
{\rm Tr}{\bf 1}=2^{[D/2]}.
\end{equation}
A valid second order operator \cite{ZBD,MS} is obtained through squaring: 
\begin{equation}
\Delta_{fermion}=[-i\gamma_{\mu}(\partial_{\mu}-igT^{a}_fA^a_{\mu})- m{\bf 1}%
][i\gamma_{\nu}(\partial_{\nu}-igT^{b}_{f}A^{b}_{\nu})-m{\bf 1}]
\end{equation}
Using (71) and (72), this may be written as 
\begin{equation}
\Delta_{fermion}=(\partial_{\mu}-igT^{a}_{f}A^{a}_{\mu})^{2}{\bf 1}+ m^{2}%
{\bf 1}-g\sigma_{\mu\nu}T^{a}_{f}F^{a}_{\mu\nu}/2.
\end{equation}
This may now be compared with (38) and one finds 
\begin{equation}
(N_{fermion})_{\mu}=-igT^{a}_{f}A^{a}_{\mu}{\bf 1},
\end{equation}
\begin{equation}
(Y_{fermion})_{\mu\nu}=-igT^{a}_{f}F^{a}_{\mu\nu}{\bf 1},
\end{equation}
\begin{equation}
X_{fermion}=-g\sigma_{\mu\nu}T^{a}_{f}F^{a}_{\mu\nu}/2.
\end{equation}
The first nonvanishing ${\cal L}^{(1)}_{fermion} (D,p)$ follows from (26)
and is 
\begin{equation}
{\cal L}^{(1)}_{fermion}(D,2)=-\frac{\hbar g^{2}}{2(4\pi)^{D/2}} \frac{%
\Gamma(2-D/2)}{m^{4-D}}\frac{2^{[D/2]}}{6}T_{f}F^a_{\mu\nu}F^a_{\mu\nu}
\end{equation}
where Tr$(T^{a}_{f}T^b_f)=T_f\delta^{ab}$ has been used. The corresponding
divergent Lagrangian is given by 
\begin{equation}
{\cal L}^{(1)}_{fermion}(D\rightarrow 4,2)=-\frac{\hbar g^{2}}
{32\pi^{2}(4-D)}\frac{4}{3}T_{f}F^{a}_{\mu\nu}F^{a}_{\mu\nu}
\end{equation}
Up to this order and as a check on our work, let us combine the results for the
divergent Lagrangians of the pure Yang-Mills (54), the scalar boson (67) 
and the Dirac fermion (80). 
\begin{equation}
{\cal L}^{(1)}_{YM}(D\rightarrow 4,2)=\frac{\hbar g^{2}}{32\pi^{2}(4-D)}
\left( \frac{11}{3}C+\frac{1}{6}T_{s}-\frac{4}{3}T_{f} \right)
F^{a}_{\mu\nu}F^{a}_{\mu\nu}.
\end{equation}
The result (81) is the one-loop divergent Lagrangian for the Yang-Mills
theory with scalar bosons and Dirac fermions and is not restricted to the
quasilocal cases (15) and (16). Except for the scalar sector, one may
compare (81) with reference \cite{AGS} which gives the $\beta$ function for
Yang-Mills with Dirac fermions. }

{\normalsize The quasilocal assumption causes ${\cal L}^{(1)}_{fermion}(D,3)$
to vanish. The next nonvanishing Lagrangian is of order $F^{4}$, 
\begin{eqnarray}
{\cal L}^{(1)}_{fermion}(D,4) =& -\frac{\hbar g^{4}}{2(4\pi)^{D/2}} \frac{%
\Gamma(4-D/2)}{m^{8-D}}2^{[D/2]}{\rm Tr} \left(
T^{a}_{f}T^{b}_{f}T^{c}_{f}T^{d}_{f} \right)  \nonumber \\
&\left(\frac{1}{18}F^a_{\mu\nu}F^b_{\nu\mu}F^c_{\rho\sigma}F^d_{\sigma\rho}- 
\frac{7}{180} F^a_{\mu\nu}F^b_{\nu\rho}F^c_{\rho\sigma}F^d_{\sigma\mu}
\right),
\end{eqnarray}
whose divergent part in 8 dimensions is 
\begin{eqnarray}
{\cal L}^{(1)}_{fermion}(D\rightarrow 8,4) =&-\frac{\hbar g^{4}}
{256\pi^{4}(8-D)}{\rm Tr} \left( T^{a}_{f}T^{b}_{f}T^{c}_{f}T^{d}_{f} \right)
\nonumber \\
& \left( \frac{8}{9}F^{a}_{\mu\nu}F^{b}_{\nu\mu}F^{c}_{\rho\sigma}
F^{d}_{\sigma\rho}-\frac{28}{45}F^{a}_{\mu\nu}F^{b}_{\nu\rho}
F^{c}_{\rho\sigma}F^{d}_{\sigma\mu} \right)
\end{eqnarray}
Finally, let us collect the results for the pure Yang-Mills (55), scalar (68)
and fermion (83) sectors and exhibit the quasilocal divergent Lagrangian in 8
dimensions. 
\[
{\cal L}^{(1)}_{YM} = \frac{\hbar g^{4}}{256\pi^{4}(8-D)} \left\{ \left[ -%
\frac{7}{48}f^{efa}f^{fgb}f^{ghc}f^{hed}  +\frac{1}{288}{\rm Tr}%
(T^{a}_{s}T^{b}_{s}T^{c}_{s}T^{d}_{s}) \right. \right. 
\]
\[
\left. -\frac{8}{9}{\rm Tr}(T^{a}_{f}T^{b}_{f}T^{c}_{f}T^{d}_{f}) \right]
F^{a}_{\mu\nu}F^{b}_{\nu\mu}F^{c}_{\rho\sigma}F^{d}_{\sigma\rho}  + \left[ 
\frac{41}{60}f^{efa}f^{fgb}f^{ghc}f^{hed} + \right. 
\]
\begin{equation}
\left. \left. \frac{1}{360}{\rm Tr}(T^{a}_{s}T^{b}_{s}T^{c}_{s}T^{d}_{s}) +%
\frac{28}{45}{\rm Tr}(T^{a}_{f}T^{b}_{f}T^{c}_{f}T^{d}_{f}) \right]
F^{a}_{\mu\nu}F^{b}_{\nu\rho}F^{c}_{\rho\sigma}F^{d}_{\sigma\mu} \right\}
\end{equation}
}

{\normalsize The ten and twelve dimensional results are too complicated to
write down and are probably not worth exposing. Nevertheless, the existence
of higher dimensional divergent Lagrangians suggests that any theoretical
model done in $D$-spacetime dimensions must incorporate the 
$F^{D/2}$-invariant (together with properly chosen ${\cal D}F$-invariants of 
the same order in the non-quasilocal case) into the bare Lagrangian. This is 
entirely in keeping with the dimensional nature of the gauge field $A$ and 
coupling $g$, namely
$$ [A] = M^{D/2 -1},\qquad [g] = M^{2-D/2}, $$
and accords perfectly with results (50) - (57). To reiterate, it is not 
enough to assume that the field theory starts off with a bare Lagrangian such 
as (39). In quantum field theory one should at the very least include extra 
powers of $F$, as encapsulated in the divergent parts (82) - (84), etc. }
\vspace{.2in}

\noindent
{\bf Acknowledgements}
\vspace{.1in}

\noindent
{\normalsize E. Rodulfo wishes to thank DOST-ESEP for financial support in
the form of a scholarship grant and the Physics Department of the University
of Tasmania for providing academic facilities.}
\vspace{.2in}

\noindent
{\bf Appendix}
\vspace{.1in}

\noindent
{\normalsize To find the functions $P(s)$, $Q(s)$ and $R(s)$, let us
preoperate the Green function equation (20) by $\int d^D\!x e^{ip \cdot
(x-x^{\prime})}$ and substitute the trial solution (21). 
$$
\int d^D\!\!x \,e^{ip \cdot (x-x^{\prime})} \left[
\partial^{2}+\!m^{2}+\!X(x^{\prime})+ 
(x-x^{\prime})_{\mu}(Y_{\mu\nu}(x^{\prime})\partial_{\nu}- 
Y^{2}_{\mu\nu}(x^{\prime})(x-x^{\prime})_{\nu}/4) \right] 
$$
$$
\times i \int_0^\infty ds \int\frac{d^D\!p^{\prime}}{(2\pi i)^{D}} 
e^{-(m^{2}+X)s-P(s)-[i(x-x^{\prime})+Q(s)] \cdot\! p^{\prime}-
p^{\prime}\cdot\! R(s)\cdot\! p^{\prime}/2} 
$$
$$
= \int d^D\!\!x \,e^{ip\cdot (x-x^{\prime})}\delta^D(x,x^{\prime}).
$$
Using the delta function representation $\int d^Dxe^{i(p-p^{\prime})\cdot
x}=(2\pi)^D  \delta^D(p-p^{\prime})$, this yields 
\[
\int_{0}^{\infty} ds \left[ -p^{2}+m^{2}+X(x^{\prime})+ 
Y_{\mu\nu}(x^{\prime})p_{\nu} \frac{\partial}{\partial p_{\mu}}+ \frac{1}{4}
Y^{2}_{\mu\nu}(x^{\prime}) \frac{\partial^{2}}{\partial p_{\mu} \partial
p_{\nu}} \right] 
\]
$$
e^{-(m^{2}+X)s-P(s)-Q(s) \cdot p-\frac{1}{2}p \cdot R(s) \cdot p} = i^{D-1}.
$$
From the antisymmetry of $Y$ and the symmetry of $R$ one can prove that 
$$
p_{\mu}Y_{\mu\nu}R_{\nu\rho}p_\rho =\frac{1}{2} p_\mu[Y,R]_{\mu\rho}p_\rho.
$$
But the complementary quasilocal condition (16) implies that $P$, $Q$ and $R$
will depend only on $Y$ and will therefore commute with $Y$. Hence, the
right-hand-side above vanishes and whenever $Y_{\mu\nu}R_{\nu\rho}$ is
encountered, we may set it to zero. Performing the $p$%
-differentiations leads to 
\[
\int_{0}^{\infty} ds \left[ m^{2}+X+ \frac{1}{4}(Q \cdot Y^{2} \cdot Q-{%
\mbox tr}Y^{2}R)+Q \cdot (-Y+ \frac{1}{2} Y^{2}R) \cdot p + \right. 
\]
$$
\left. \frac{1}{2} p\cdot (-2 +\frac{1}{2} R Y^{2} R)\cdot p \right] \times
e^{(m^{2}+X)s-P(s)-Q(s) \cdot p - \frac{1}{2} p \cdot R(s) \cdot p} =
i^{D-1},
$$
which is clearly integrable provided $P$, $Q$ and $R$ satisfy the first
order differential equations, 
$$
\frac{\partial}{\partial s} P(s) =\frac{1}{4}[Q(s) \cdot Y^{2} \cdot Q(s) - 
{\rm tr} Y^{2}R(s)],
$$
$$
\frac{\partial}{\partial s} Q(s) =Q(s) \cdot \left[ -Y+ \frac{1}{2}
Y^{2}R(s) \right],
$$
$$
\frac{\partial}{\partial s} R(s) = -2+ \frac{1}{2} R(s) Y^{2} R(s).
$$
By this means, 
$$
\left. e^{-(m^{2}+X)s-P(s)-Q(s) \cdot p- p \cdot R(s) \cdot p/2}
\right|_{0}^{\infty} = -i^{D-1}
$$
The differential equations for $P,Q,R$ subject to the boundary
condition (22) give us the following solutions: 
$$
P(s) = -{\rm tr}[\ln \sec(iYs)]/2
$$
$$
Q(s)=0
$$
$$
R(s) = 2iY^{-1} \tan(iYs)
$$
}

{\normalsize Performing the $p$-integration in the trial solution (21) using
the formula 
$$
\int d^{D}\!p\,e^{-P-Q \cdot p -\frac{1}{2} p \cdot R \cdot p} = i^{D}
\pi^{D/2} s^{-D/2} e^{-P+ \frac{1}{2} Q \cdot R^{-1} \cdot Q - \frac{1}{2} 
{\rm tr} \ln(-R/2s)}
$$
then setting $x=x^{\prime}$, one finds that the Green function evaluated 
at zero separation is 
$$
\langle h(x)h(x)\rangle =\frac{\hbar}{(4\pi)^{D/2}} \int_{0}^{\infty}
ds\,s^{-D/2} e^{-(m^{2}+X)s-{\rm tr} \ln[(iYs)^{-1} \sin(iYs)]/2}.
$$
Substituting this into (13) and integrating with respect to $X$ while
remembering the limiting value (12), one finds \cite{BD,DRM} 
$$
{\cal L}^{(1)} = \frac{\hbar}{2(4\pi)^{D/2}} \int_{0}^{\infty} \frac{ds}{%
s^{1+D/2}} e^{-m^{2}s} {\rm Tr} \left\{ e^{-Xs-{\rm tr} \ln[(iYs)^{-1}
\sin(iYs)]/2}-e^{-X(0)s} \right\}.
$$
Last but not least, the Taylor expansion 
$$
\ln\left[\frac{\sin(z)}{z}\right]=\sum_{n=1}^{\infty}\frac{(-1)^{n}2^{2n}
B_{2n}}{2n (2n)!}z^{2n},
$$
where $B_{n}$ are the Bernoulli numbers, allows one to rewrite the last
integral in the form (23) which is suitable for the purpose of obtaining 
${\cal L}^{(1)}$ to any specific order in the fields $X$ and $Y$. }

\end{document}